\documentstyle[12pt]{article}

\begin{document}
\pagestyle{headings}
{\bf \ \ } \\
\begin{flushright} {\bf ISU-IAP.Th95-01,\ Irkutsk\\
                         hep-ph/9504261} \end{flushright}
\vspace{2cm}
\begin{center}
{\Large \bf Another look at the angular distributions of the
$\gamma\gamma\to\pi \pi$ reactions.}\\[2.5cm]

{\Large A.E.Kaloshin$^{a}$, V.M.Persikov$^{a}$ and
V.V.Serebryakov$^{b}$}

\end{center}

\begin{flushleft}
{\sl$^{a}$ Institute of Applied Physics,
Irkutsk State University,
664003 Irkutsk, Russia\\
$^{b}$ Institute for Mathematics, 630090
Novosibirsk, Russia\\[1cm] }
\end{flushleft}
{\large Abstract.}
\vspace{0.5cm}

We analyze the existing data on the angular distributions of the
$\gamma\gamma\to \pi^+ \pi^- ,\ \pi^0 \pi^0$  reactions with using of
the unitary model for helicity 2 amplitude. The purpose is to obtain
the D--wave parameters and S--wave cross section.  We obtain from
experiment in the first time the values of $\alpha + \beta$ for sum
of the electric and magnetic pion's polarizabilities.  We found the
S-wave cross sections  much smaller as compared with previous similar
analysis.  Comparison of the $\gamma\gamma\to \pi^+ \pi^-$ and
$\gamma\gamma\to \pi^0 \pi^0$ data gives an indication for a marked I
= 2, J = 2 contribution in region of $f_2(1270)$.\\[0.5cm]

\newpage
{\Large 1. Introduction.}
\vspace{0.3cm}

Nowadays there exist few experiments on the angular distributions of
the $\gamma\gamma\to \pi \pi$ reactions: with charged pions from CELLO
\cite{CELLO} and MARK-II \cite{MARK} detectors and
 with neutral pions from Crystal Ball \cite{CB}.
The most evident feature of these data is the resonance peak
$f_2(1270)$ interfering  with smooth background.
But the most interesting physically is
the S-wave cross section and this interest
is related with long-standing problem of scalar mesons
whose properties and spectrum are looking as mysterious.
Note that
there is another not investigated earlier aspect of low energy physics
-- that's a possibility of experimental study of the sum
$(\alpha + \beta)^{\pi}$ of electric and magnetic pion's
polarizabilities.
The only experimental information on these values is the restriction
for  $(\alpha + \beta)^{\pi^+}$ from experiments on nuclei \cite{Antipov}.
\footnote{Let's recall that the sum $\alpha + \beta$ plays the role
of the D-wave threshold structure constant in helicity 2 amplitude,
similar to electromagnetic radius in a formfactor.
As for the S-wave parameter $\alpha - \beta$, the existing
near--threshold data $\gamma\gamma\to \pi \pi$ allow to get the
first information on it, see \cite{KS-93}.}
We found that even existing $\gamma\gamma\to \pi \pi$  experiments
allow to have much more exact information on $\alpha+\beta$,
the main problem here is an investigation of systematical
inaccuracy in such an analysis. The matter is that one needs to
extrapolate formulae with obtained parameters from
$s \sim \ 1 \ GeV^2$ to point  s = 0.

The question is appearing how to use the existing experimental data on
angular distributions to obtain the physical information. It's not
so evident question, so let's say few words on this matter.

In ideal situation of high quality data the most preferable
way is the model--independent partial wave analysis.
But really it's not so profitable with present data. Such an
attempt was made in \cite{Harjes} and it was found that
results are very indefinite
because of incomplete solid angle ( $\mid cos \Theta \mid < 0.6$)
of detector and presence of higher spin waves
in $\gamma\gamma\to \pi^+ \pi^-$ from QED ($\pi$ - exchange)
mechanism.

The second possibility, realized in \cite{CELLO},
consists in using of the model for dominating helicity 2
amplitude $T_{+-}$, where the main physical effects are more or less
clear. This way allowed to obtain from angular distributions
the most interesting S-wave cross section with low statistical
errors. Evidently this way needs the accurate modelling
of the main contribution.

The third way \cite{MP} consists in using of the model expressions
for both helicity amplitudes  $T_{+-}$ and $T_{++}$.
The main problem here is related with the modelling of
I = 0 S-wave and is generated  by the contradictory situation and
insufficient understanding of hadron dynamics in this sector.
In this case one should use the multi--channel approach
and it needs the additional model assumptions
with an additional uncertainty.

Our starting point is the desire to look at the S-wave
contribution in the maximal model--independent way. So we choose
the second way a'\ la CELLO with modeling of the main contribution
and in the following we have all pluses and minuses of this choice.
As compared with \cite{CELLO} we use another (more developed)
model for dominating amplitude and include into consideration
also the $\gamma\gamma\to \pi^0 \pi^0 $ reaction.\\[0.3cm]

{\Large 2. Model for helicity 2 amplitude.} \vspace{0.2cm}

We shall use the model \cite{KS-86} for helicity 2 amplitude
which has the following properties:
\begin{enumerate}
\item  The lowest wave with  J = 2, I = 0  satisfies
the one--channel unitary condition. As it was shown e.g. in
\cite{KS-88}, the inelastic effects in the process
$\gamma\gamma\to f_2(1270) \to \pi \pi$  are rather small.
\item  The J = 2, I = 2 wave does not take into account the
final state interaction effects and contains only Born
contributions.
\item The model satisfies the low energy theorem
\cite{LET} not only on the level of Tompson limit but with
accounting of first structural correction too.
\item It satisfies also the unsubtructed dispersion relation
at fixed t \cite{AG}.
\end{enumerate}

Helicity amplitudes in CMS are defined in a standard manner.
The cross section $\gamma\gamma\rightarrow\pi^+\pi^-$ is
(for neutral pions there appears an additional factor 1/2!):
\begin{equation}
\frac{d \sigma}{d cos \Theta} = \frac{\rho(s)}{64 \pi s}  \
\{\ \mid T_{++}\mid ^2 + \mid T_{+-}\mid ^2  \} \\[0.5cm]
\end{equation}

Here $\rho(s) = (1-4\mu^2/s)^{1/2}$ ,\ $\mu= m_{\pi}$.
Let's pass over to reduced helicity amplitudes M which are free of
kinematical singularities and zeroes \cite{AG}.
\begin{equation}
T_{++} = s M_{++}, \ \ \ \ \ T_{+-} = (tu-\mu^4) M_{+-}
\label{Reduc}
\end{equation}
The reduced helicity 2 amplitudes \cite{KS-86} with
minimal modification are looking as :
\begin{eqnarray}
M^C_{+-} = \frac{1}{\sqrt{3}D_f(s)}\ \left[\ C^0 + H^0_V(s)\ \right] +
W^C(s,t) \nonumber\\
M^N_{+-} = \frac{1}{\sqrt{3}D_f(s)}\ \left[\ C^0 + H^0_V(s)\ \right] +
W^N(s,t)
\label{amp}
\end{eqnarray}
Here and below C means "Charged" and N  "Neutral".\\
The background contributions have the form:
\begin{eqnarray}
W^C(s,t) &=& W^{\pi}(s,t) +
\sum_{V=\rho,b_1,a_1} Z^V \left(\frac{1}{m_V^2-t} + \frac{1}{m_V^2-u} -
\frac{1}{m_V^2}\right)
- 5 \frac{Z^{a_2}}{m^2_{a_2}+s}
+ a^C \frac{s_{eff}}{s + s_{eff}}\nonumber\\
W^N(s,t) &=&
\sum_{V=\rho,\omega,h_1,b_1} Z^V \left(\frac{1}{m_V^2-t} + \frac{1}{m_V^2-u}
- \frac{1}{m_V^2}\right)
+ a^N \frac{s_{eff}}{s + s_{eff}}
\label{back}
\end{eqnarray}
Here $W^{\pi}(s,t)$ is the contribution of $\pi$--exchange (QED),
$Z^R = g^2_{R\pi\gamma}/4$.
Besides the known resonance exchanges we allow the presence of
contributions which are not taken into account exactly -- these are
the terms proportional to arbitrary constants $a^C, a^N$.
Firstly, the existing experimental information on decays
$R \to \pi\gamma$ \ allows the marked freedom in the cross--
contributions.
Secondly, the introduced terms should absorb in a some way
and another physical effect: the modification of $\pi$--exchange
due to presence of some off-shell formfactor.
 The existence of parameter $s_{eff}$
accounts additional arbitrariness in extrapolation
to point $s = 0$.

$D_f^{-1}(s)$ is the propagator of $f_2(1270)$--meson with
accounting of the finite width corrections.
\begin{equation}
D_f(s) = (m_f^2 - s)\cdot(1+Re H^{\prime}(m_f^2)) - Re H(s) +
Re H(m_f^2) -  i m_f \Gamma_f(s)
\end{equation}
$\Gamma_f(s)$ is chosen in a standard way with accounting of
centrifugal barrier.
\begin{equation}
\Gamma_f(s) = \Gamma_f\cdot \left[\frac{q_{\pi}(s)}{q_{\pi}(m_f^2)}
\right]^5 \cdot \frac{D_2(r q_{\pi}(m_f^2)}{D_2(r q_{\pi}(s)},\ \ \ \
D_2(x) = 9 + 3 x^2 + x^4
\label{gamma}
\end{equation}
The parameter r defines the D--wave scattering length
 $a^0_2$.

Function $H_V^0 (s)$ in (\ref{amp}) is the rescattering contribution.
\begin{equation}
 H^{0}_{V}(s) = \frac{s}{\pi}\int \frac{ds'}{s'(s'-s)}
m_f \Gamma_f(s') V^{0}(s') \ ,
\label{Resc}
\end{equation}
where  $V^{0}(s)$  is projection of background contributions
(\ref{back}) onto J = 0, I = 0 state.
The function $R(s)=C^0+Re \ H_V^0(s)$ at resonance point
defines the two--photon decay width  $f_2 \rightarrow \gamma \gamma $ .
\footnote{In real case of the marked background, interfering
with $f_2(1270)$, the question what we should call by the two--photon
width becomes transparently
unambiguous at current experimental accuracy.
The considerable deviation of values
$f_2 \rightarrow \gamma \gamma $ from different groups data
may be explained in part by different definitions. }
\begin{equation}
 R(m_f^2)=\frac{\sqrt{3} \cdot 2^5 \cdot 5\pi}{m_f^3}
\sqrt{\Gamma_f \cdot \Gamma(f \rightarrow \gamma \gamma) \cdot
BR(f \rightarrow \pi \pi)}
\end{equation}
The pion polarizabilities are defined at the point s = 0
(e.g. \cite{KS-86}) and we have from (\ref{amp}) :
\begin{eqnarray}
\frac{(\alpha + \beta)^C}{2\mu}&=& \frac{C^0}{\sqrt{3}D_f(0)} +
\sum_{V=\rho,b_1,a_1} \frac{Z^V}{m_V^2}
- 5 \frac{Z^{a_2}}{m^2_{a_2}}
+ a^C \nonumber\\
\frac{(\alpha + \beta)^N}{2\mu}&=& \frac{C^0}{\sqrt{3}D_f(0)} +
\sum_{V=\rho,\omega,h_1,b_1} \frac{Z^V}{m_V^2}
+ a^N
\end{eqnarray}
The amplitudes (\ref{amp}) contain three essential parameters
$C^0, a^C, a^N$. Instead of them we can use another set of parameters
(linear combinations of the first set)
$R^0=C^0 + Re \ H_V^0(m_f^2), \ (\alpha + \beta)^C,\
(\alpha + \beta)^N$ , which is preferable from our point of view.
Passing to this set is rather simple.\\[0.2cm]

{\Large 3. Discussion of the CELLO  analysis.} \vspace{0.3cm}

As it was said in Introduction, we shall analyze an angular
distributions in the same semi--model manner  \cite{CELLO},
which is looking as the most attractive. For helicity 2
amplitude we use the model expression with few free parameters
but the S--wave contribution is extracted independently in
every energy bin. So it's necessary to say few words about the
results of \cite{CELLO} where the data of  \cite{CELLO} and \cite{MARK}
$\gamma\gamma\to \pi^+ \pi^-$ were analyzed.
They used for helicity 2 amplitude a
rather simplified expression, satisfying the one--channel unitary
condition.
In symbolic form:
\begin{equation}
T_{+-} = T_{+-}^{QED} + B.-W.\ f_2(1270)
\label{symb}
\end{equation}
But if to look on results \cite{CELLO} more carefully,
there are rather unexpected statements.
\begin{enumerate}
\item CELLO and MARK--II data need the significant
damping of the QED contribution, in 1 GeV region the value of this
damping is few times in any case.
\item There was found the S--wave contribution of unexpectedly big scale:
$\sigma^S (\gamma \gamma \rightarrow \pi^+ \pi^- , \mid cos \Theta \mid
< 0.6 )\sim 60 - 80 $ \ nbarn at $\sqrt{s} = 0.8 - 1.0 $ \ GeV.
It differs significantly from natural scale $\sim 10 $ nbarn,
appearing both at the simple estimates  \cite{KS-86}, based on
polarizability,
and at the extrapolation of near--threshold analysis \cite{KS-93}
to this region.
\end{enumerate}
It turns out (see for more details \cite{KP}) that
these results are based on the specific assumption about
form of the Breit--Wigner contribution in (\ref{symb}).
It was taken as:
\begin{equation}
\frac{d \sigma^{BW}}{d \mid cos \Theta \mid} = 40 \pi \cdot
\frac{m_f^2}{s} \ \cdot
\frac{\Gamma_{\gamma \gamma} \ \Gamma_f(s) \ BR(f \rightarrow \pi^+ \pi^-)}
{\mid m_f^2 - s -i m_f \Gamma_f(s) \mid^2} \cdot
\mid Y_{22}(cos \Theta) \mid^2 \\[0.5cm]
\label{Ansatz}
\end{equation}

The expression (\ref{Ansatz}) is valid for narrow resonance, when the
chosen
s--dependence in nominator is unessential, but for rather broad  $f_2$
such choice is looking
as arbitrary. This assumption is very essential for analysis
\cite{CELLO},
so let's look at consequence of (\ref{Ansatz}).
$\Gamma_f(s)$ was chosen as :
\footnote{In fact in \cite{CELLO} it was put f(s)=1,
the introduction of any decreasing with s factor makes all the
problems even stronger.}
\begin{equation}
\Gamma_f(s)= \Gamma_f \cdot \left [ \ \frac{q(s)}{q(m_f^2)} \ \right ]^5
\cdot
\frac{m_f}{\sqrt{s}} \cdot f^2(s)
\end{equation}
f(s) is so called factor of centrifugal barrier.
Returning from (\ref{Ansatz}) to the lowest partial wave
one can find :
\begin{equation}
(M_{+-})^{J=2,I=0} = R(s)/(m_f^2 - s -i m_f \Gamma_f(s)),\ \ \ \
R(s) = \frac{const}{s} \ f(s)
\label{R_H}
\end{equation}
The main difference between the model (\ref{amp}) and \cite{CELLO}
consists in behavior of R(s).
In the model \cite{KS-86} the effective "coupling constant"
$R(s)=C^0+Re \ H^0_V(s)$ is defined by rescattering effect and is much
more smooth function in vicinity of resonance. Moreover, the
appearance of the pole in  R(s) (\ref{R_H}) breaks the low energy
theorem  requirements.
In an analysis the value $R(m_f^2)$
is fixed well by data. However in case of (\ref{R_H}) the function
R(s) grows essentially below the resonance and it gives too big
D--wave cross section exceeding the experiment. Just this fact
leads to necessity to damp  the QED contribution and in the end
gives all above mentioned results. The repetition of analysis
with another model \cite{KS-86} gives other results:

a)  For description of experimental data both on total
cross section and angular distributions  it's not necessary
any additional QED damping at the same or even better
quality of description.

b) The extracted S--wave contribution
$ \sigma^S ( \mid cos \Theta \mid < 0.6 ) $
is much less than was it found in  CELLO analysis and does not
contradict to results of threshold analysis  \cite{KS-93}.\\[0.3cm]

{\Large 4. Analysis of angular distributions.} \vspace{0.3cm}

Our helicity 2 amplitudes (\ref{amp}) contain three parameters:
$R^0=C^0 + Re \ H_V^0(m_f^2),\\ (\alpha + \beta)^C,\
(\alpha + \beta)^N$.
We found at numerical investigation, that corresponding cross section of
$\gamma \gamma \rightarrow \pi^+ \pi^-$  depends very weakly
on the $(\alpha + \beta)^N$  and
$\gamma \gamma \rightarrow \pi^0 \pi^0$ practically does not depend
on $(\alpha + \beta)^C$.
So for a single reaction we can use the two--parameter expression
fixing the alien polarizability somewhere in theoretically
expected region. One can change it in a few times without any
marked changing in results.
Recall that different low energy quantum field models \cite{Din,VO}
give rather close values :
$(\alpha + \beta)^{\pi^+} \simeq  0.20 $,
$(\alpha + \beta)^{\pi^0}  \simeq  1.20 $ in units of $10^{-42} \ cm^3$ .
\footnote{We use the units system  $e^2=4 \pi \alpha$, where
the values of polarizabilities differ by factor $4 \pi$  as compared
with the system $e^2=\alpha$. Below we shall use the
units  $10^{-42} \ cm^3$  for polarizabilities
not indicating them.}
Some greater values are predicted by the dispersion sum rules
\cite{Petr} and two--loop calculations chiral model calculations
\cite{chm2} (see Table 2 for more details).

So we shall describe the experimental angular distributions in the
following way:
\begin{equation}
\frac{d \sigma}{d cos \Theta} = a_S \  +  \
\frac{\rho(s)}{64 \pi s}  \
(tu-\mu^4)^2\cdot \mid M_{+-}(s,t)\mid ^2
\label{fit}
\end{equation}
and use the model (\ref{amp}) with two free parameter for
helicity 2 amplitude for every reaction.

We found also another source of uncertainty
related with the parameter r in centrifugal barrier (\ref{gamma})
or in other words with D--wave scattering length. As for this
parameter, it practically does not change the $\chi^2$
value in analysis influencing ,however, on the extracted
low energy parameters.

So the sources of systematical inaccuracy in our analysis
are the uncertainties in parameter r (D-wave scattering length)
and  in model for background (\ref{back})
interfering with resonance.

Let us restrict ourselves in analysis by the region $E < 1.5 \ GeV$ ,
and  take $m_{f_2}$ ,  $\Gamma_{f_2}$ from PDG Tables \cite{PDG}.
At the first step let's fix the value r in (\ref{gamma})
by $r=5.5\ GeV^{-1} $, it corresponds to the standard scattering
length value $a^0_2 = 1.7\cdot 10^{-3}$  in units of pion mass.
This scattering length was obtained from experiment
\cite{Petersen} $a^0_2 = (1.7\pm 0.3) \cdot 10^{-3}$  and it was used
to fix counterterms in the chiral model loop calculations \cite{GL,chm2}.

The results of our analysis at fixed $r=5.5\ GeV^{-1} $
are shown in Table 1. There are indicated few variants corresponding
to different forms of background contribution. Let us note few
facts seen from this Table.
\begin{itemize}
\item The quality of description in all cases is satisfactory
with exception ,may be, the CELLO data. In this case , however,
quality of description is  better than in analysis
\cite{CELLO} of the same data. Indeed $\chi^2/NDF = 81.4/53$
in \cite{CELLO} and 69/51 in our analysis.
\item  In all variants of description  the sum $\alpha + \beta$
is defined with very small statistical error.
\footnote{Recall that the only experimental information on
the sum $\alpha + \beta$   is the following:
$(\alpha+\beta)^C = 1.8\pm 3.9\pm 3.1$  \cite{Antipov}.}
It means that  $d\sigma/dc$  in vicinity of $f_2(1270)$
( D--wave parameters are defined in main by resonance
vicinity due to big statistical weight) is extremely sensitive
to  value of background contribution.
\item Both parameters $\alpha+\beta$ for $\pi^+$ and $\pi^0$
are lying in expected regions.
\item  The S--wave contributions near 1 GeV are much less than in
\cite{CELLO} in all cases and correspond on the scale to
results of near--threshold analysis \cite{KS-93} and to old
estimates \cite{KS-86} based on the value $\alpha-\beta$.
The typical scale for the obtained S-wave cross section is
$\sim$ \  10  \ nbarn.

\end{itemize}

Our results for polarizabilities are summarized in Table 2
in comparison with existing predictions for these values.
Recall that it was obtained with standard parameters of
$f_2(1270)$, generally accepted form of $\Gamma_f(s)$
and standard scattering length.\\
\vspace{0.2cm}
\underline{On the S--wave contributions.}\\
Together with D--wave parameters, shown in Table 1, we obtain
the S--wave contributions in every energy bin (\ref{fit}).
Our results for them, corresponding to variant 2 of Table 1,
are shown in Figures 1--3. The other variants of Table 1 have
qualitatively  the same behavior. Note that we allow parameter $a_S$
in (\ref{fit})
to be negative in a fit. As a result we see that the S--wave
contributions turn out much less than in analysis \cite{CELLO}.
In case of CELLO data one can see some indications
on the scalar
meson $\epsilon(1300)$ production in this process but we can
see from Fig. 1--3 that resonance picture is not so transparent.
For numerical estimate let's consider the CELLO  S--wave, assuming
the resonance production with mass 1200 MeV and width 300 MeV.
Then the extracted cross section of Fig. 1 corresponds to
$\Gamma(\epsilon \to \gamma \gamma)\cdot BR(\epsilon \to \pi \pi)
\sim  3.6 \ KeV$ --see the curve. For MARK-II and Crystal Ball data
there is no evident resonance--like picture and S-wave is
less than in CELLO case.

One can see from these Figures one exclusive case: that's for
MARK--II data in the region
$E \leq 0.9 \ GeV$ , where the S-wave cross section is formally
negative. It happens in all variants 1--4 of Table 1.
This circumstance practically does not influence on
the D--wave parameters since they are defined mainly by resonance
region. Besides, this "negative cross sections"
have rather small value as compared with total cross sections.

It's not so difficult to understand the origin of this effect.
Sure the cross section
$\gamma \gamma \rightarrow \pi^+ \pi^-$
differs from QED one because of final state interaction
effects. But at the standard form of $\pi \pi$--interaction
(i.e. smooth extrapolation of $\pi \pi$ phase shift
from resonance to threshold with the positive scattering length)
the D--wave cross section exceeds the QED one in this region.
\footnote{Introducing of some formfactor to QED vertex does not
help here if you do not break the low energy theorem. Besides,
this degree of freedom is absorbed rather well by our
"effective cross--exchange" . The same problem but in much more
sharp form was ,evidently, in analysis \cite{CELLO},
which has been lead to necessity of additional
"damping factor" breaking the low energy theorem
at  the level of structure corrections.}
However in the region  $E \leq 0.9 \ GeV$
the MARK--II data in contrast to CELLO are below then the
QED contribution -- see Fig.4 . Naturally with given type of analysis
(model for helicity 2 amplitude) there is no place for S--wave
and these contributions will be negative.
Let us note that and previous experiments
$\gamma \gamma \rightarrow \pi^+ \pi^-$ (measuring of integral
cross section) differ from each other in this aspect: some of them
obtain the cross section higher than QED curve, and some lower.
\\[0.3cm]
\underline{On two--photon width of $f_2(1270)$.}\\
At more detailed looking at Table 1 one can see the  disagreement
in two--photon coupling constant $R^0$ between
$\gamma\gamma\to \pi^+ \pi^-$  and   $\gamma\gamma\to \pi^0 \pi^0$
experiments. To demonstrate it
let us list the two--photon width, corresponding
to the variant 2 of
Table 1, with statistical errors only (systematical ones are much less).
\footnote{Here we shall accept
$\Gamma(f_2 \to \gamma \gamma)\sim (R^0)^2$ for simplicity. In fact
we would not like to discuss in this paper what is most correct
definition  for decay width in this case.}
\begin{eqnarray}
CELLO:\ \ \Gamma(f_2 \to \gamma \gamma) = 2.95\pm0.13 \ KeV \nonumber \\
MARK-II:\ \ \Gamma(f_2 \to \gamma \gamma) = 2.84\pm0.18 \ KeV \nonumber \\
Crystal Ball:\ \ \Gamma(f_2 \to \gamma \gamma) = 3.70\pm0.22 \ KeV
\end{eqnarray}

Even taking into account that this discrepancy is related with
different experiments, we see that the difference may reach to
three standard deviations and it should be considered
seriously. So let's consider the possible physical reasons for it.
\footnote{We mentioned in above that the D--wave parameters
are defined mainly by the resonance vicinity only. So we think this
discrepancy is not related with problem of negative S--wave,
if it really exists.}
\begin{itemize}
\item First of all this suggests that one should take into
account the effects of final state interaction in J = 2, I = 2
wave too. There exist some experimental information on this
phase shift $\delta^2_2$ : it is slow and negative in wide region
(see, i.g.\cite{Alekseeva} ). Indeed, we made such an attempt and found
that this effect reduces the difference. But its influence is too
small: roughly speaking we shall have the difference about
two standard deviations instead of three.
\item Perhaps the data indicate on deviation of the $f_2(1270)$
parameters from generally accepted. Considering the mass and total
width as free parameters we  found the best  $\chi^2$
at  $m_{f_2}=1.28$ GeV and $\Gamma_{f_2}=230$ MeV.
But it does not reduce the difference in  $R^0$.
Besides, the problem of negative S--wave contributions
becomes much more sharp. So this possibility seems to be
unreasonable.
\item  One more possible reason: if in the $\pi$--exchange
the above mentioned off--shell formfactor plays the essential role.
In the lowest partial wave this effect is absorbed by
our "effective cross--exchange" (we checked it in few examples)
but the higher spin waves  $J > 2$  in (\ref{amp})
do not contain this effect. As a result of corresponding
calculations we came to conclusion that this effect works
in the opposite direction: any damping of higher spin waves
$J > 2$ leads to stronger contradiction for the two--photon
coupling constant $R^0$.
\end{itemize}

We came to opinion that most probable reason of this
disagreement in  $f_2(1270)\ \gamma \gamma$ coupling
is related with some non--standard D--wave dynamics with I = 2.
Let's recall that the observed in the processes
$\gamma\gamma\to\rho\rho$ anomaly near threshold
( $\sigma(\gamma\gamma\to\rho^0\rho^0 ) \gg
\sigma(\gamma\gamma\to\rho^+\rho^-)$ )
is interpreted almost unambiguously as an exotic resonance I = 2
production (see e.g. discussion in \cite{Achasov}).
But this effect can not be considered in framework of
one--channel approach and is far away of purposes of present work.

Finally, what will be changed in results if to vary the
parameter r in the centrifugal barrier ? Let it changes in interval
$4.0 < r < 6.0\ GeV^{-1}$, it corresponds to D--wave
scattering length between $ 0.6 \cdot 10^{-3}$ and  $ 2.2 \cdot 10^{-3}$.
We shall have for polarizabilities :
\begin{eqnarray}
CELLO:\ \ (\alpha+\beta)^C = 0.37\pm0.08(stat.)\pm0.10(syst.)
   \nonumber \\
MARK-II: \ \ (\alpha+\beta)^C = 0.23\pm0.09(stat.)\pm0.12(syst.)
\nonumber  \\
Crystal Ball:\ \ (\alpha+\beta)^N = 1.40\pm0.10(stat.)\pm0.26(syst.)
\end{eqnarray}

As for two--photon coupling constant, it is very stable and
the S--wave contributions will have practically the same
behavior.\\[0.3cm]

{\Large 5. Conclusions.} \vspace{0.3cm}

Thus, we performed the semi--model analysis a' la CELLO \cite{CELLO}
 of existing data on the angular distributions of
$\gamma \gamma \rightarrow \pi \pi$  for both reactions.
In contrast to \cite{CELLO} we used another
model for the helicity 2 amplitude \cite{KS-86} which does not break the
low energy theorem requirements.

We came to natural conclusion that such kind of analysis needs
the very accurate modelling of dominating amplitude.
The essential difference between our results and \cite{CELLO}
tells that one should utilize in the model an information on
$\pi \pi$ - interaction in rather wide region.
The control of threshold parameters both of hadron and electromagnetic
amplitudes is very useful here.

Another our observation: the angular distributions in resonance
vicinity are very sensitive (especially for neutral pions)
to background value. This degree of freedom is absolutely necessary
for data describing. We gave  a physical sense to these
degrees of freedom, relating them with pion's polarizabilities,
but it's not a necessary step.

There are few facts which are convinced ourselves in correctness
of our approach:
\begin{itemize}
\item  In our analysis the S--wave cross sections in region
of 1 GeV and below have the typical scale about 10 nbarn,
which corresponds to reasonable values of difference
of polarizabilities \cite{KS-93,KS-86}.
\item There is no necessity for introducing of any additional
damping of QED contributions at least in the first approximation.
Even if here is a problem, it is much more soft as compared with
\cite{CELLO}.
\item The obtained values for polarizabilities sum both for
 $\pi^+$ and $\pi^0$ in any variant do not contradict
to theoretical predictions (see Table 2).
\end{itemize}

We came to conclusion that using of the model  \cite{KS-86}
for helicity 2 amplitude leads to rather agreed picture
at least on the level of large contributions. We met some
contradictions in our analysis too, but so to speak on the
next level. The contradictions appear either with rather small
S--wave contributions or at comparison of different experiments.
The most serious one is the difference in
the two--photon coupling constant of $f_2(1270)$  from
$\gamma\gamma\to \pi^+ \pi^-$  and   $\gamma\gamma\to \pi^0 \pi^0$
experiments. In our opinion it tells  about new physical
effect not included into standard description.

As for "negative" cross sections in MARK-II data
(see Fig.2): the appearance of this effect is related
with chosen form of analysis. But we think that here exists also the pure
experimental problem of more exact calibration of the
measured cross section --
see Fig. 4 for illustration. We have in mind the location
of experimental cross section relatively the $\pi$--exchange
contribution's curve -- sure that's much more delicate question
than the cross section measuring.

We suppose that the physical results of the performed analysis
are the following:
\begin{itemize}
\item  We obtained from two--photon experiments in the first time the
sum of polarizabilities both for $\pi^+$ and $\pi^0$. The existing
data allow to extract the background contributions interfering
with resonance $f_2(1270)$ with very small statistical errors.
Thought there exists some freedom at the extrapolation to point
s = 0, it is not so big as one could think from the beginning.
It's surprising, but due to existing of the "amplifier"  $f_2(1270)$,
there are even better conditions for obtaining  the D--wave parameter
$\alpha + \beta$  from data than for the S--wave one $\alpha - \beta$.
\item The obtained S--wave cross sections are rather small parts
of the total cross sections, their scale is about 10 nbarn.
There exists some resonance--like enhancement near 1.3 GeV of rather
small amplitude in CELLO case. The obtained S--wave in region of $E \leq 1$
does not contradict to results of near--threshold analysis \cite {KS-93}.
\item  We observe the statistically meaningful difference
between $\gamma \gamma \rightarrow \pi^+ \pi^-$ and
$\gamma \gamma \rightarrow \pi^0 \pi^0$
experiments in value of $f_2(1270) \gamma\gamma$ coupling.
The most probable reason is the existing of non--standard dynamics
in  I = 2, J = 2  wave.
\end{itemize}

 As for comparison with results of Morgan and Pennington \cite{MP},
it's difficult to say unambiguously does our S--wave contradict to
their result or not. They have few solutions (with accounting of
$\Gamma^{++}_{f_2 \gamma \gamma}$ coupling or not), their preferable
solution has resonance--like behavior of the I=0 S--wave cross section
with much bigger value. This solution has the sizeable
$\Gamma^{++}_{f_2 \gamma \gamma}$ coupling. We here restricted
ourselves by assumption $\Gamma^{++}_{f_2 \gamma \gamma} = 0$,
as it was made in \cite{CELLO}. We didn't meet serious contradictions
with this assumption at least in the first approximation. The
inclusion of this coupling into consideration needs the essential
hypothesis because of interference effects.\\[0.3cm]

{\Large Acknowledgments.} \\[0.2cm]
We are grateful to J.K. Bienlein  for kind sending
the information on Crystal Ball data.
A.E.K. would like to thank M.Scadron for fruitful discussion
and  S.Cooper for useful remarks. The research described in this
publication was made possible in part by Grant No NN7000 from the
International Science Foundation.

\newpage
Table 1.\\[0.3cm]

{\small
\noindent
\begin{tabular}{|c|c|c|c|c|}   \hline
&1&2&3&4    \\ \hline
&$\rho\ + \ \omega$,&$\rho\ + \ \omega$,&All resonances,
&All resonances,    \\ &$s_{eff}=\infty$&$s_{eff}=1.69 \ GeV^2$&$
s_{eff}=\infty$&$s_{eff}=1.69 \ GeV^2$ \\ \hline \hline
&&&&\\
CELLO&$R^0=0.290\pm 0.009$&$0.291\pm0.009$&$0.290\pm0.009$&
$0.291\pm0.009$  \\
$\gamma \gamma \rightarrow \pi^+ \pi^-$&$(\alpha+\beta)^C=0.41\pm0.04$
&$0.40\pm0.08$ &$0.41\pm0.04$&$0.42\pm0.08$\\
$0.8\leq E \leq 1.5$&$\chi^2/NDF=69/51$ &69.2/51 & 69/51 & 69/51   \\
&&&& \\ \hline
&&&& \\
MARK-II&$R^0=0.282\pm 0.014$&$0.286\pm0.013$&$0.283\pm0.014$&
$0.285\pm0.013$  \\
$\gamma \gamma \rightarrow \pi^+ \pi^-$&
$(\alpha+\beta)^C=0.32\pm0.05$&$0.23\pm0.09$
&$0.32\pm0.05$&$0.25\pm0.09$\\
$0.6\leq E \leq 1.5$&$\chi^2/NDF=23.2/48$
&23.6/48&23.2/48& 23.4/48   \\
&&&& \\ \hline
&&&&\\
CB&$R^0=0.332\pm 0.013$&$0.326\pm0.013$&$0.331\pm0.013$&
$0.324\pm0.013$  \\
$\gamma \gamma \rightarrow \pi^0 \pi^0$&$(\alpha+\beta)^N=1.24\pm0.06$&
$1.56\pm0.10$ &$1.32\pm0.06$&$1.62\pm0.10$\\
$0.85\leq E \leq1.45$&$\chi^2/NDF=44.0/47$
&42.0/47&43.6/47& 41.6/47    \\
&&&& \\ \hline
\end{tabular}                        }

\vspace{1cm}
Table 2.\\[0.3cm]
{\small
\noindent
\begin{tabular}{|c|c|c|c|c|c|c|}   \hline
&Present work&\multicolumn{2}{c|}{Chiral models}&Superconduct.
&Quark--virton&Dispersion \\  \cline{3-4}
&&One loop&Two loops&quark model&model & sum rules  \\
&&\cite{chm1}&\cite{chm2}&\cite{VO}&\cite{Din}&\cite{Petr}\\ \hline
$(\alpha+\beta)^C$&$0.41\pm0.08\pm0.01$&0&--&0.2&0.2&$0.42\pm0.05$ \\
&(CELLO)&&&&&     \\
&$0.28\pm0.09\pm0.05$&&&&&$$ \\
&(MARK--II)&&&&&    \\ \hline
$(\alpha+\beta)^N$&$1.43\pm0.10\pm0.20$&0&$1.45\pm0.38$&1.20&1.2&$
1.61\pm0.08$   \\
&(Crystal Ball)&&&&&\\ \hline
\end{tabular}                       }
\newpage
\begin{center}
\Large Tables captions:
\end{center}
\begin{itemize}
\item {\bf Table 1}\ \
Best--fit D--wave parameters at fixed value $r=5.5\ GeV^{-1} $,
it corresponds to scattering length  $a^0_2 = 1.7\cdot 10^{-3}$.
Different variants 1--4 correspond to different forms of
background contribution (\ref{back}). $R^0$ (the two--photon
coupling constant of $f_2(1270)$)
in units of $GeV^{-2}$,
$\alpha + \beta$ in units of $10^{-42} cm^3$, $e^2 = 4 \pi \alpha$.
\item {\bf Table 2}\ \
Comparison of obtained values for the sum of polarizabilities
at $r=5.5\ GeV^{-1} $ ($a^0_2 = 1.7\cdot 10^{-3}$)
with existing theoretical predictions.

\end{itemize}  \hspace{1cm}

\begin{center}
\Large Figures captions:
\end{center}
\begin{itemize}
\item {\bf Figure 1 }\ \
Best--fit S--wave cross section $\mid cos \Theta \mid < 0.6$
from CELLO data \cite{CELLO}
at fixed value $r=5.5\ GeV^{-1} $. The points with central box are
result of analysis \cite{CELLO} of the same data. For illustration
there is shown the curve corresponding to scalar meson production
with $M = 1200 \ MeV,\ \Gamma = 300\ MeV$  and \\
$\Gamma(\epsilon \to \gamma \gamma)\cdot BR(\epsilon \to \pi \pi)
 = 3.6 \ KeV$.
\item {\bf Figure 2 }\ \
The same for MARK--II data \cite{MARK}.
\item {\bf Figure 3 }\ \
The same for Crystal Ball data \cite{CB}, $\mid cos \Theta \mid < 0.7$.
\item {\bf Figure 4 }\ \
Integral cross sections of CELLO and MARK--II below 1 GeV,\\
$\mid cos \Theta \mid < 0.6$, in comparison with $\pi$--exchange
helicity 2 contribution (curve).
\end{itemize}

\newpage
\begin{thebibliography}{88}
\bibitem{CELLO} CELLO Coll. H.Behrend  et al.,\ Z. Phys. {\bf C 56}
, 381 (1992).
\bibitem{MARK} MARK-II Coll. J.Boyer et al., \ Phys. Rev. {\bf D 42},
1350 (1990).
\bibitem{CB} Crystal Ball Coll.
J.K.Bienlein,  Report at 9th Int.Workshop
on Photon--Photon Coll.,La Jolla, 23--26 March 1992 ;
Preprint DESY-92-083 and J.K. Bienlein (private communication).
\bibitem{Antipov} Yu.M.Antipov et al.,\  Z. Phys. {\bf C 26}, 495 (1985).
\bibitem{KS-93} A.E.Kaloshin and V.V.Serebryakov, \ Yad. Fiz. {\bf 56},
114 (1993).
\bibitem{Harjes} J.Harjes, \ DESY Internal Report DESY-FCE-91-01, 1991.
\bibitem{MP} D.Morgan and M.R.Pennington,\  Z. Phys. {\bf C 48}, 623 (1990).
\bibitem{KS-86} A.E.Kaloshin and V.V.Serebryakov,\ Z. Phys. {\bf C 32},
279 (1986).
\bibitem{KS-88} A.E.Kaloshin and V.V.Serebryakov,\ Yad. Fiz. {\bf 47},
179 (1988).
\bibitem{LET} F.T.Low,\ Phys. Rev. {\bf 96}, 1428 (1954).
\bibitem{AG} H.Abarbanel and M.L.Goldberger,\ Phys. Rev. {\bf 165},
1594 (1968).
\bibitem{KP} A.E Kaloshin and V.M.Persikov, \ Yad. Fiz. {\bf 56}, 203 (1993).
\bibitem{Din} M.Dineihan et al., \ JETP Lett. {\bf 35}, 443 (1982).
\bibitem{VO} M.K.Volkov and A.A.Osipov, \ Yad. Fiz. {\bf 41}, 1027 (1985).
\bibitem{Petr} V.A.Petrun'kin,\ Phys. Elem. Part. Atom. Nuclei {\bf 12},
692 (1981).\\
A.I.L'vov and V.A.Petrun'kin, \ Preprint FIAN  303 (1985).
\bibitem{chm2}  S.Bellucci, J.Gasser and M.E.Sainio, \
Nucl. Phys. {\bf B 423}, 80 (1994).
\bibitem{PDG} Particle Data Group,\ Phys. Rev. {\bf D 50}, 1173 (1994).
\bibitem{Petersen} J.L.Petersen,\ Phys. Rep. {\bf C2} 155 (1971) .\\
CERN Report 77-04 (1977).
\bibitem{GL} J.Gasser and H.Leutwyller,\ Ann. of Phys. {\bf 158}, 142 (1984).
\bibitem{Alekseeva}  E.A.Alekseeva et al.,\ JETP \ {\bf 55}, 591 (1982).
\bibitem{Achasov} N.N.Achasov and G.N.Shestakov,\ Uspehi Fiz. Nauk
{\bf 161}, 53 (1991).
\bibitem{chm1} M.K.Volkov and V.N.Pervushin,\ Nuovo Cimento {\bf 27A},
277 (1975). \\
J.Bijnens and F.Cornet,\ Nucl. Phys. {\bf B296}, 557 (1988).\\
J.F.Donoghue, B.R.Holstein and Y.C.Lin,\ Phys. Rev. {\bf D 37}, 2423 (1988).
\end{thebibliography}
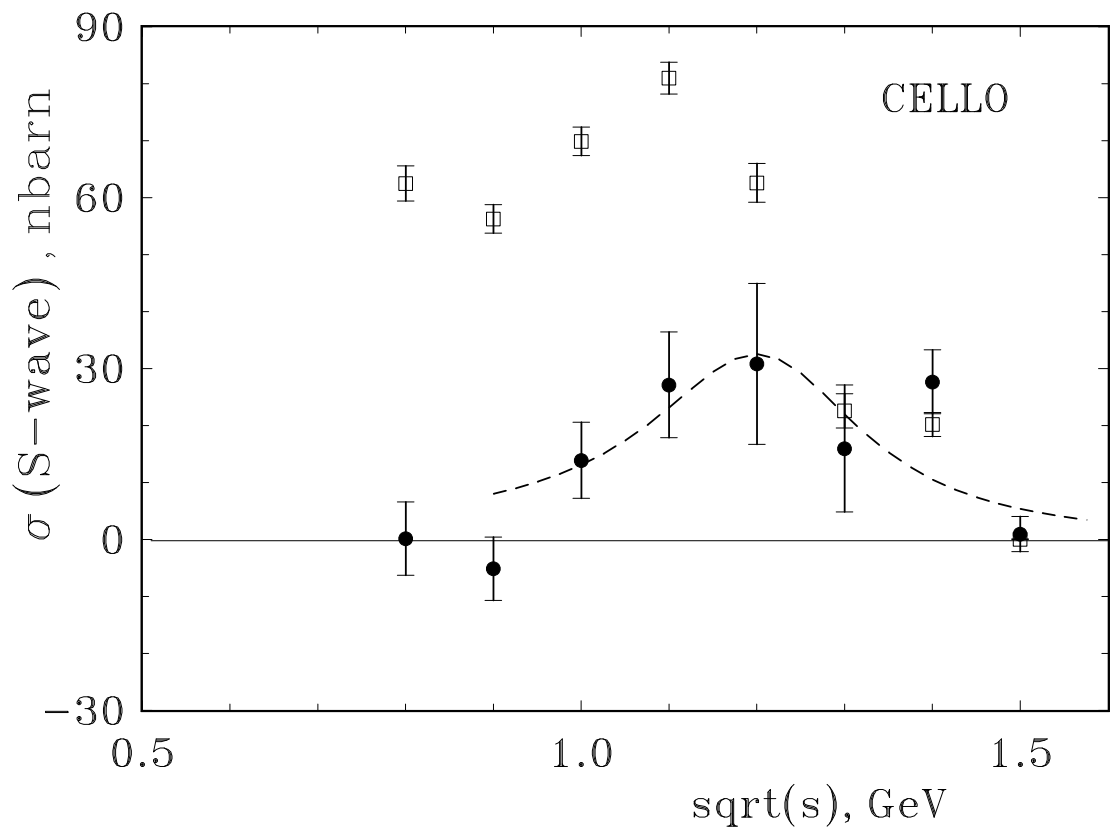
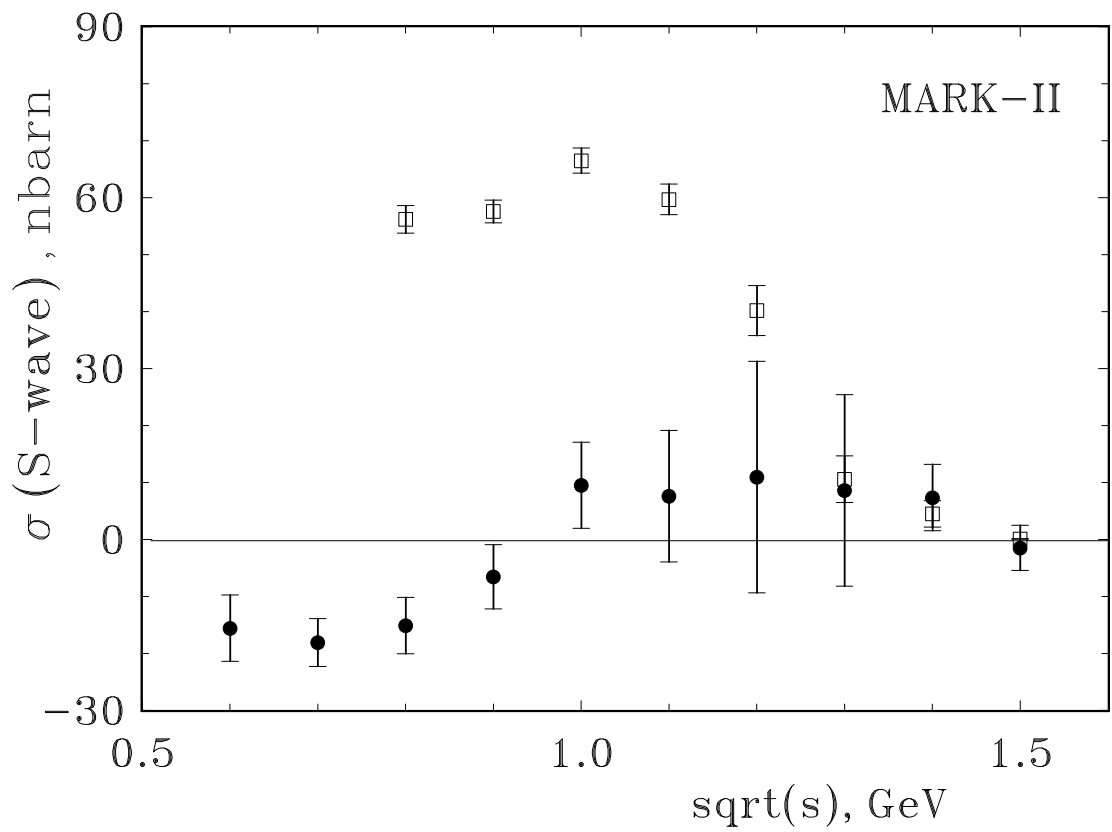
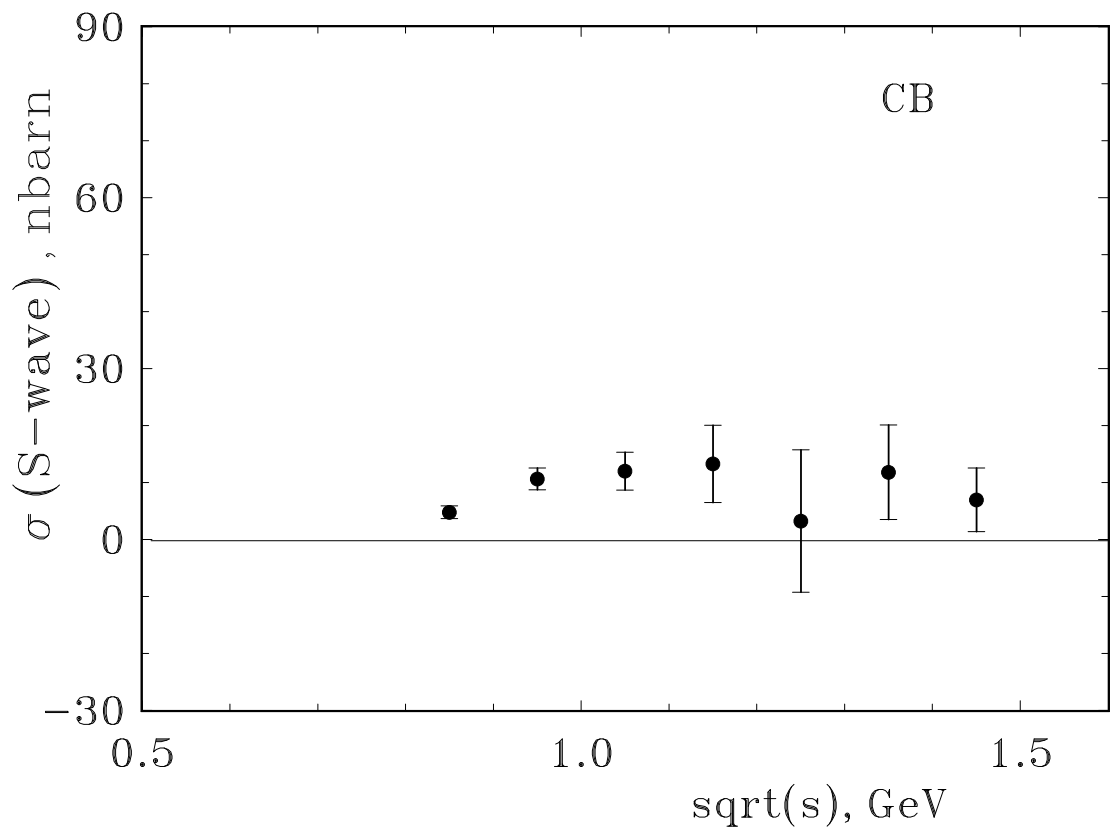
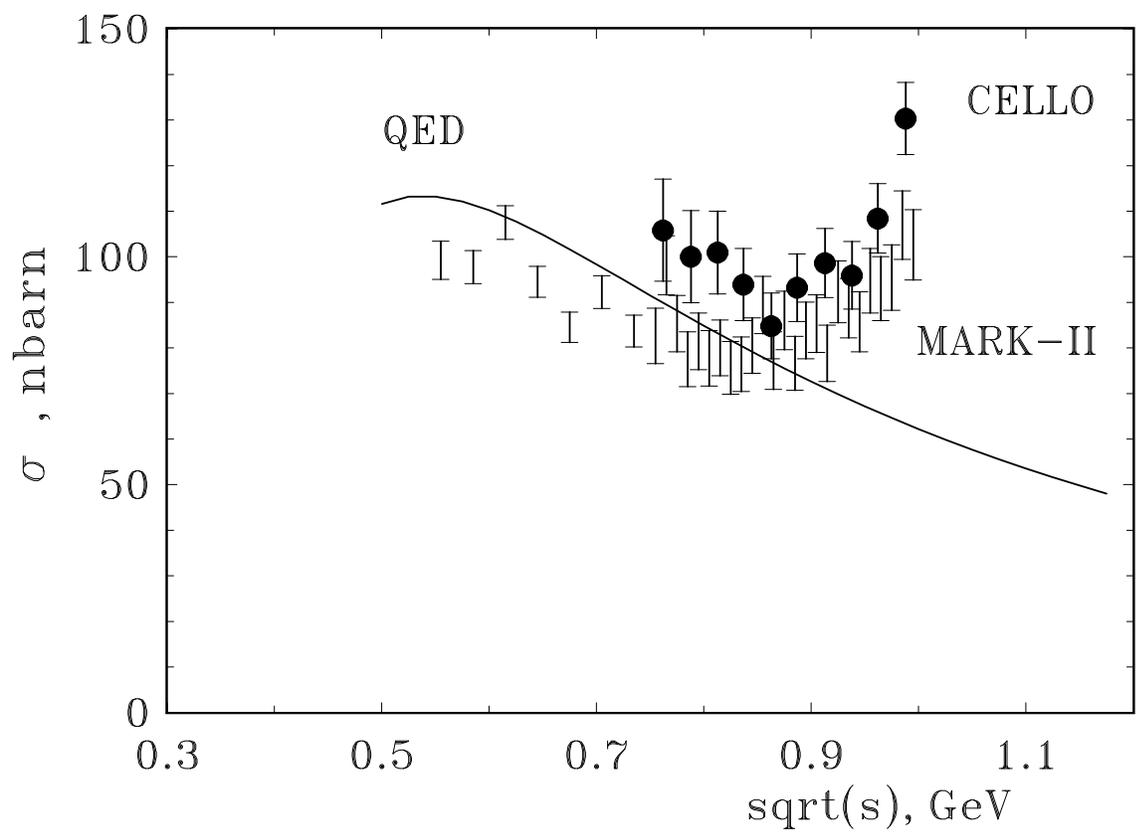
\end{document}